\documentclass[11pt,a4paper]{iopart}

\usepackage{graphicx,color,fancybox}
\usepackage{units}
\usepackage{yfonts}
\usepackage{paralist}
\usepackage{hyperref}
\expandafter\let\csname equation*\endcsname\relax

\expandafter\let\csname endequation*\endcsname\relax

\usepackage{amsmath,amssymb,amsfonts}
\usepackage[title,titletoc,toc]{appendix}
\usepackage{tensor,mathrsfs}
\usepackage{listings}
\usepackage{mdframed}

\makeatletter
\newcommand{\mainmatter}{%
  \setcounter{footnote}{0}%
  \patchcmd{\@makefntext}{\fnsymbol}{\arabic}{}{}%
  \patchcmd{\@thefnmark}{\fnsymbol}{\arabic}{}{}%
  \def\@makefnmark{\textsuperscript{\arabic{footnote}}}%
}
\makeatother

\makeatletter
\def\footnoterule{\kern-3\p@
  \hrule \@width 2in \kern 2.6\p@} %
\makeatother

\begin{document}

\newcommand{\exd}{\mathrm{d}}
\newcommand{\reals}{\mathbb{R}}
\newcommand{\prin}{P(M,G)}
\newcommand{\prinmap}{P\xrightarrow{\pi} M}
\newcommand{\parmder}{\frac{\exd}{\exd t}}
\newcommand{\thth}{\textsuperscript{th}}
\newcommand{\integers}{\mathbb{Z}}
\newcommand{\pe}{\pi_{PE}}

\newcommand{\CC}[1]{\textcolor{blue}{[Casey: #1]}}

\title{Relating Gauge Gravity to String Theory Through Obstruction}
\author{Casey Cartwright$^{1,2}$\footnote{Corresponding author}, Alex Flournoy$^{2}$\footnote{Deceased - November 3, 2023}}%
\address{$^1$ Institute for Theoretical Physics, Utrecht University, Princetonplein 5, 3584 CC Utrecht, The Netherlands}

\address{$^2$ Colorado School of Mines, Golden, Colorado 80401, USA}

\ead{c.c.cartwright@uu.nl}
\ead{aflourno@mines.edu}

\vspace{10pt}
\begin{indented}
\item[]\date{}
\end{indented}

\begin{abstract}
In this article we provide a more detailed account of the geometry and topology of the composite bundle formalism introduced by Tresguerres in Phys. Rev.
D 66 (2002) 064025~\cite{Tresguerres2002} to accommodate gravitation as a gauge theory. In the first half of the article we identify a global structure required by the composite construction which not only exposes how the ordinary frame and tangent bundle expected in general relativity arise but provides the link or connection between these ordinary bundles and the gauge bundles of the composite bundle construction. In the second half of the article we discuss implications of this method of constructing gravity as a fiber bundle for the global structure of spacetime. We find that the underlying manifold of the composite bundle construction is expected to admit, not only a spin structure but also a string structure. As a consequence of our work we are able to extend past work on global structures of physically reasonable spacetime manifolds. It has been shown that in four spacetime dimensions, that if an oriented, Lorentzian, four dimensional manifold is stably casual that it is parallizable, and hence admits a spin structure which allows for chiral spinors. We may now add to this that such a manifold also admits a string structure.    
\end{abstract}


\submitto{\CQG~\footnote{This is the Accepted Manuscript version of an article accepted for publication in Classical and Quantum Gravity. IOP Publishing Ltd is not responsible for any errors or omissions in this version of the manuscript or any version derived from it.  The Version of Record is available online at \href{https://doi.org/10.1088/1361-6382/ada966}{doi:10.1088/1361-6382/ada966}}}
\maketitle

\section{Introduction}\setcounter{footnote}{0}
The history of attempts to formulate gravity as a gauge theory began shortly after the seminal work of Yang and Mills~\cite{yang}. In the following year Utiyama made the first attempt to obtain Einstein's field equations from a gauge principle by localizing the Lorentz symmetry of flat spacetime~\cite{ryoyu}. His efforts elucidated the complications of gauging external (or spacetime) symmetry groups, but fell short of providing a fully consistent derivation of general relativity from a gauge principle. Among its shortcomings were the ad hoc introduction of local Lorentz frames and the absence of a conserved energy-momentum tensor acting as a source of curvature. In 1961 Kibble, identifying spacetime translations as the generator of the energy-momentum current, extended Utiyama's analysis to a gauge theory based on the full Poincar\'e group~\cite{kibble}. Kibble's construction, often referred to as Poincar\'e gauge theory, reproduced the field equations, in particular realizing curvature sourced by energy-momentum, however it also required an additional torsion term sourced by the spin-angular momentum of fields and did not provide a consistent interpretation of the role played by coframes. In the fifty years since Kibble's work there has been no shortage of attempts to understand how gravitation is realized from a gauge principle~\cite{transgauge1,Ivanenko1983,Sardanashvily2005,lord86,Tresguerres2002,Hehlaffine,poinc2groupquangrav,higherTPG,DCpro}.

While much of the work on gravitation as a gauge theory has proceeded in terms of action functionals, less effort has aimed at finding an underlying construction in terms of a principal fiber bundle~\footnote{Such a description exits for $(2+1)$-dimensional gravity as a Chern-Simons theory as shown by Witten~\cite{Witten:1988hc}}. This despite the necessity of the bundle formalism for addressing gauge theories in topologically nontrivial spacetimes. An immediate obstacle to any such construction based on Poincar\'e symmetry is clear. The Lorentz subgroup acts on the tangent-space indices of fields at a given point in spacetime and hence can be realized as fiberwise or ``vertical" transformations which leave the base-point unchanged. However elements of the translation subgroup necessarily move between points in the base. This presents a difficulty in using the usual interpretation of gauge transformations as vertical fiber automorphisms. It was suggested by Lord that these could be accommodated by allowing ``horizontal" components of the fiber action~\cite{lord86}. This idea was given a more complete realization in the work of Tresguerres who utilized the formalism of composite bundles~\cite{Tresguerres2002}. Investigating these matters carefully, we find a restriction on spacetime imposed as a sufficient condition to obtain a consistent composite bundle construction. This leads to the requirement that the manifold allow a Lorentzian signature metric, that it be consistent with other known gauge theories i.e.\ that it be orientable, and that it admit a spin structure which allows for chiral fermions. However a further finding came as something of a surprise. The consistency of the composite bundle formulation seems to imply that the underlying spacetime also admit a string structure. This of course is the relevant restriction on the target spaces of perturbative string models.

The outline of this paper is as follows. We begin with a review of the standard fiber bundle formalism as it applies to gauge theory in order to establish notation. We then review the composite bundle formalism and in particular how it is applied to gauge theories of gravitation. Utilizing a consistency condition from composite fiber bundles, we then investigate implications for the structure of the base spacetime and enumerate our findings. We conclude with comments on the significance of these findings and directions for future work.    

\section{Bundles}
\label{sec:fiberbundles}
\subsection{Ordinary fiber bundles} 
\begin{figure}[ht]
 \begin{center}
  \includegraphics[width=3.75in]{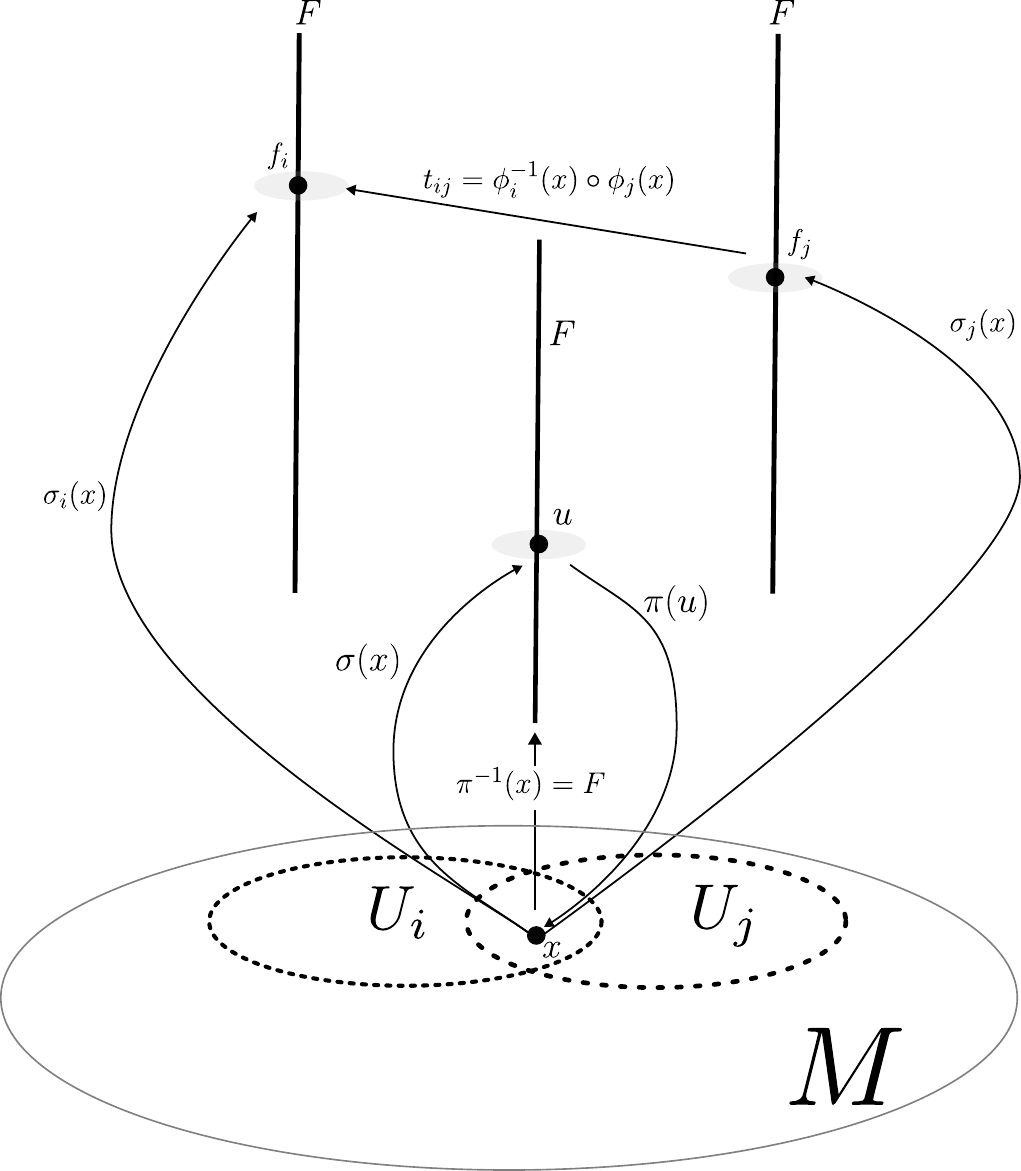}
 \end{center}
\caption{\label{fig:fiberbun}A fiber bundle is shown. The base $M$ is a differentiable manifold which we cover by open sets, where we have displayed two overlapping open neighborhoods in the cover. For each point of the neighborhoods, and so the whole manifold, we attach a fiber space $F$. The collection of all of the fiber spaces over all of the open neighborhoods is what we call $E$ the total space.}
\end{figure}

In what follows we follow closely the expositions in Nakahara and Frankel~\cite{nakahara,frankel}. A differentiable fiber bundle (see \fref{fig:fiberbun} for a pictorial representation) denoted by $E\xrightarrow{\pi}M$ or $E(M,F,G,\pi)$ consists of the following data,
\begin{enumerate}
 \item Differentiable manifolds $E$,$M$ and $F$, called the total, base and fiber space respectively.
 \item A surjection $\pi :E\rightarrow M$ called the projection.
 \item A Lie group $G$ called the structure group such that $G$ has left action on the fiber.
 \item An open cover $\{U_i\}$ of the base with diffeomorphism $\phi_i:U_i \times F \rightarrow \pi^{-1}(U_i)$ such that $\pi \circ \phi_i(p,f)=p\in M$.
 \item On every non empty overlapping set of neighborhoods $U_i\cap U_j$ we require a $G$ valued transition function $t_{ij}=\phi_i\circ\phi_j^{-1}$ such that $\phi_i=t_{ij}\phi_j$.
\end{enumerate}
A fiber bundle is in general an extension of the Cartesian product between two spaces $X$ and $Y$, $X\times Y\equiv \{(x,y)|x\in X, \hspace{.1cm} y\in Y\}$. A non-trivial fiber bundle then cannot be realized as a standard Cartesian product globally. However locally we do have a trivialization of a bundle $E(M,F)$ as $M\times F $ so it is typical to work with local pieces defined on subsets of the space $M$. An important concept is that of a \textit{section} of a fiber bundle. A section $\sigma:M\rightarrow E$ is a smooth mapping such that $\pi\circ\sigma(p)=id_{M}$. If a section is defined only on a chart $U_i$ then we say $\sigma_i$ is a local section. Sections are assignments of points in the base space with points in the total space $E$ and induce mappings of vectors in the base space to vectors in the total space. \\

\textit{Principal bundles ---} Of particular relevance for gauge theory are principal fiber bundles wherein the fiber space is taken to be the structure group $G$ itself denoted $P\xrightarrow{\pi} G$ or $P(M,G)$. In addition to the left group action of $G$ on the fibers we can also define a right action $R_g$ such that if $u\in P$ we have $R_gu=ug$. As an example take the principal bundle as $P(\mathbb{R}^{1,3},U(1))$. The group operation is multiplication and an arbitrary element of $U(1)$ can be expressed as $e^{i\theta}$. $\mathbb{R}^{1,3}$ is a trivial base space so we need only one coordinate chart $(U,\varphi=x^{\mu})$. We can write the local trivialization of a point $u\in P$ as $\phi^{-1}(u)=(p,e)$ where $p\in\mathbb{R}^{1,3}$ and $e\in U(1)$ is the identity element of the group. This is known as the canonical local trivialization. Given two trivializations $\phi,\phi'$ there exists an element of $G$ such that $\phi'=\phi g=\phi(ug)$. Therefore we have,
\begin{equation}
 \phi'(p,e^{i\theta})=\phi(p,e) e^{i \theta}=\phi(p,ee^{i\theta})=\phi(p,e^{i\theta}).
\end{equation}
By repeated application of the right action we move through the fiber $G$. This leads us to define the whole fiber as $\pi^{-1}(p)=\{ug|g\in G\}$. This property will hold in any principal bundle. Naturally we will always choose to represent our sections as right multiplication of the canonical local trivialization. An example of a section of a principal bundle $P(M,G)$ for $p\in M$, $e\in G$ is given by $\sigma(p)=\phi(p,e)=u\in P$. \\

\textit{Associated bundles ---} For a principal bundle $P(M,G)$ we define a vector space $V$ such that $G$ has left action on $V$ under a representation $\rho:G\rightarrow GL(n,K)$, with $K$ the field underlying $V$. If we assign an action on the product space $P\times V$ as,
\begin{equation}
 (u,\nu)\rightarrow (ug,\rho(g^{-1})\nu),
\end{equation}
the associated vector bundle is the equivalence class of points where we identify all elements of the form $(u,\nu)\sim (ug,\rho(g^{-1})\nu)$. We denote the vector bundle associated~\footnote{Matter fields are often accommodated by the introduction of a vector bundle associated with $P(M,G)$. In terms of the associated bundles a matter field is a section $\xi:M\rightarrow E$ of the bundle $E=P\times_G V$ or is simply an element of the space $C(P,V)$~\cite{bleeker}.} to $P$ by $E=P\times_G V$ or $E(M,F,G,\pi_E,P)$. The equivalent notion of the space of equivariant vector valued functions on $P$ denoted by $C(P,V)$ is much simpler to work with. If we take $\tau\in C(P,V)$ the condition of equivariance is $\tau(ug)=\rho(g^{-1})\tau(u)$, exactly the condition by which we quotient $P\times V$ to construct $E=P\times_G V$~\cite{bleeker}.  \\

\textit{Sections and triviality ---}Indeed a bundle will not always admit a global section, a mapping defined for the whole space rather then some subset of the cover. In fact the triviality of a principal bundle is reflected in whether or not it admits a global section. For vector bundles on the other hand which always admit a global zero section, their triviality is indicated by the presence of an everywhere non-vanishing global section. Otherwise the triviality of a vector bundle can be assessed by investigating global sections of its associated principal bundle. If the associated principal bundle has non-vanishing global sections then so does the vector bundle. 

\subsection{Composite bundles}
The fiber bundle formalism provides an underpinning for gauge theories that is particularly useful on topologically nontrivial spaces. The preceding analysis can be applied to any gauge theory of internal symmetry groups. However for external or spacetime symmetry groups, especially those expected to be relevant for gravitation, the standard fiber bundle formalism must be generalized. In~\ref{app:fiber} we collect some additional information about standard fiber bundles we can be referenced if the reader is interested in further contrasting a standard fiber bundle to the composite construction. 

Poincar\'e gauge theory is based on the group $ISO(1,3)=SO(1,3)\rtimes\reals^{1,3}$, i.e.\ the semi-direct product of the Lorentz group and the translations. The connection splits into a direct sum of two components~\cite{Kobayashi}, the spin connection $\omega$ and the coframe field $\theta$. The spin connection arises from the Lorentz symmetry and the coframe field from the translational symmetry. As a result there are two conserved quantities, the energy-momentum and spin-angular momentum currents, which arise from variations of the action with respect to the gauge degrees of freedom. The energy-momentum current couples to the curvature of the Lorentz connection and the spin-angular momentum current couples to the curvature of the translational connection. Both connections transform as proper connections,
\begin{equation}
 \mathcal{A}'=g^{-1}\mathcal{A}g+g^{-1}\exd g,
\end{equation}
under Poincar\'e transformations. At first glance, torsion aside, this appears to be a satisfactory gauge theory containing Einstein's general relativity. However there are two significant criticisms of this approach. First the dual coframe has one contravariant (upper) Lorentz index, thus the dual frame transforms as a contravariant vector under Lorentz rotations. This can be determined by recalling that for $\Lambda\in SO(1,3)$ the Minkowski metric is invariant $\tensor{\Lambda}{^i_a}\tensor{\Lambda}{^j_b}\eta_{ij}=\eta_{ab}$, hence
\begin{equation}
 ds^2=\hat{\theta}^i\hat{\theta}^j\eta_{ij}=\hat{\theta}^a\hat{\theta}^b\tensor{\Lambda}{^i_a}\tensor{\Lambda}{^j_b}\eta_{ij}=\hat{\theta}^a\hat{\theta}^b\eta_{ab}.
\end{equation}
This is not the case in Poincar\'e gauge theory where the coframe is a connection and transforms as such. This is a problem with the standard approach of Poincar\'e gauge theory. Either the coframe field is not a connection or the standard approach is not sensitive enough to detect the proper transformation properties.    

\begin{figure}[ht]
 \begin{center}
  \includegraphics[width=0.8\textwidth]{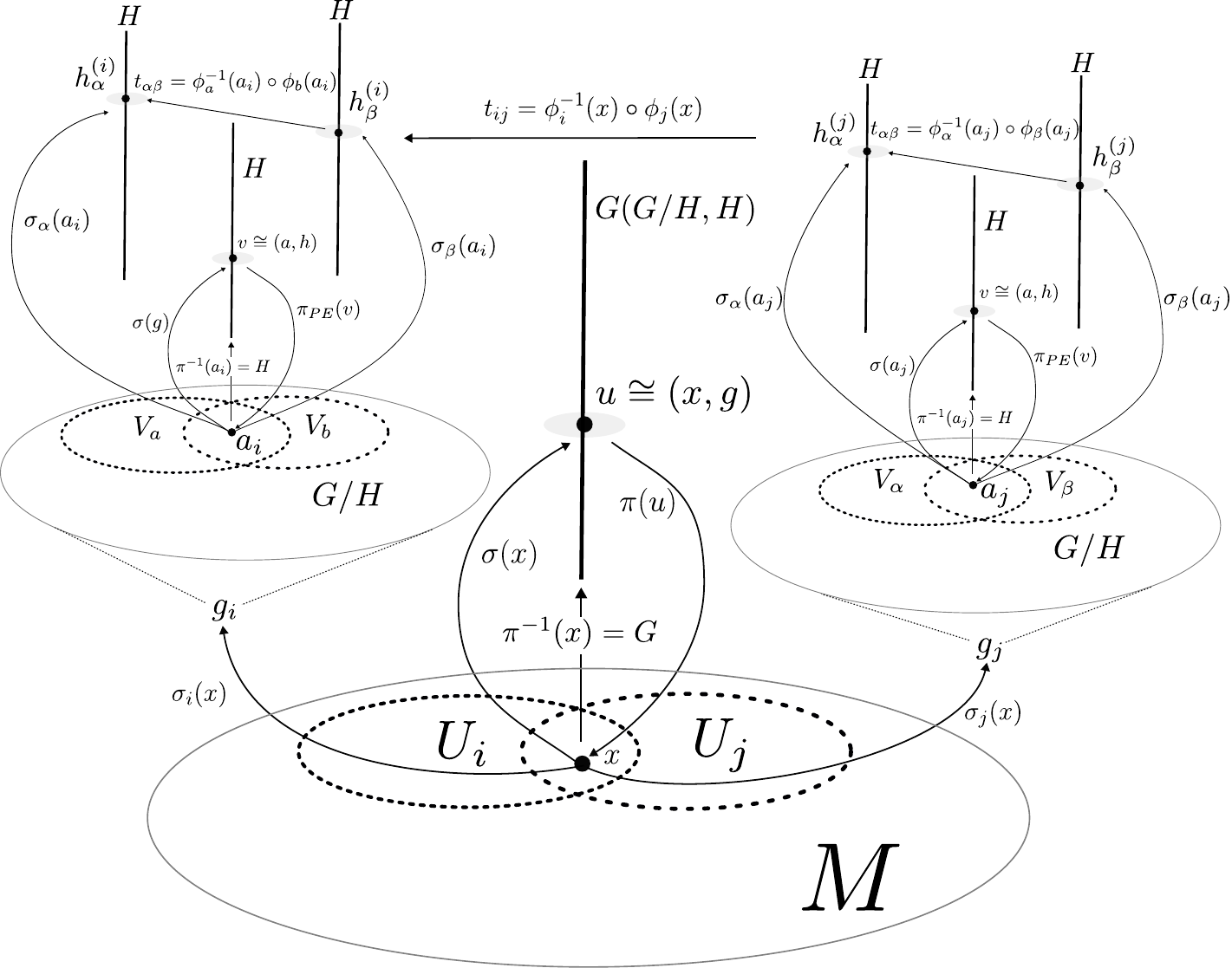}
 \end{center}
\caption{\label{fig:compbun}A depiction of a composite bundle is shown. Like ordinary fiber bundles there is a base space $M$ and fiber spaces $G$ for each point of $M$. A point $u\in P$ in the total space created by the base $M$ and fiber space $G$ is decomposed locally as $u=(x,g)\in M\times G$. However for a composite bundle the element $g\in G$ can be further decomposed $g\in G(G/H,H)$. The total space of the composite bundle is then locally $E\cong P/H\times H$ and can be decomposed further as $(x,a,h)\in M\times G/H\times H$.}
\end{figure}

The second criticism concerns the actions of the symmetry groups. In general relativity the symmetry groups are external isometry groups, i.e.\ they act on spacetime. In the previous section we worked with gauge transformations that were defined as vertical bundle automorphisms, or in other words transformations that did not move between points in the base spacetime. This was noticed by Lord in~\cite{lord86}, where he proposed that spacetime gauge theory be based on the bundle $P(G/H,H)$ with $H$ taken to be the Lorentz group, $G$ taken to be the Poincar\'e group and the verticality condition $f\circ\pi=\pi$ be relaxed. Without the verticality condition gauge transformations could induce transformations on the base spacetime. The translational component of Poincar\'e gauge theory acts to move between points in the base space, hence dropping the verticality condition seems to be a step in the right direction. However without a translational fiber space the theory lacked a translational connection. The composite bundle theory of gravitation (detailed below) introduced by Tresguerres alleviates these problems by considering a principal $G$ bundle over $M$, $P(M,G)$, to be split into a chain of bundles $P\xrightarrow{\pi_{PE}}E \xrightarrow{\pi_{EM}}M$ (\fref{fig:compbun})~\cite{Tresguerres2002}.

A consequence of using the composite scheme in the gravitational context is that the tetrad defined from the coframe can be identified as a nonlinear translational connection. This is closely related to the work of Coleman, Wess and Zumino~\cite{Coleman1969} on nonlinear sigma models. In fact the composite bundle theory of gravitation is a geometric realization of nonlinearly realized symmetry groups. Julve et al. demonstrated the connection between gravitational gauge theory and nonlinearly realized gauge symmetries~\cite{Julve1996}  and Tresguerres and Tiemblo explored the connection in the context of composite bundle formulations of gravitation~\cite{Tiemblo:2004suh}.

In what follows we present a short exposition of the composite bundle approach introduced by Tresguerres. We will then discuss a crucial aspect of the composite formalism that we believe has not been addressed in the literature. This will lead us to consider the topological restrictions arising in composite bundle formulations of gravitation.\\

\textit{Composite bundles ---} Much of the analysis of composite bundles is very similar to the internal case developed in \sref{sec:fiberbundles}. However now we will have two bundles over the base space $M$ which will be related. The idea is to work with the principle bundle $P(M,G(G/H,H))$ where $G(G/H,H)$ itself denotes a principal fiber bundle with $G/H$ as the base space and $H$ as the fiber space. We can take sections as before only now there are three possible choices: one for each sector ($\sigma_{ME}:M\rightarrow E$ and $\sigma_{EP}:E\rightarrow P$), and one for the total space ($\sigma:M\rightarrow P$). Consistency requires that $\sigma=\sigma_{EP}\circ\sigma_{ME}$. 
Sections can be written as before in terms of the local canonical trivialization, 
\begin{subequations}
\begin{align}
\sigma_{ME}(x)&=\phi_{ME}(p,e_{G/H})a(\xi) \\
\sigma_{EP}&=\phi_{EP}(u_E,e_H)h, \label{eq:compsections}
\end{align}
\end{subequations}
where we have the identity elements $e_{G/H}\in G/H$ and $e_H\in H$ and the elements $u_E\in E$, $a\in G/H$ and $h\in H$. Formally we write $u_E$ for the input to the section $\sigma_{EP}:E\rightarrow P$. However we have in mind that we are using a local section. Locally the bundle $E\rightarrow M$ can seen as $M\times G/H$ and so we can specify a point $u_E\in E$ in \Eref{eq:compsections} as $(p,\xi)$.   

Gauge transformations in the total bundle $P$ respect the conditions we imposed in the previous section, i.e.\ they commute with the right action and are vertical with respect to the fiber. However in the sector $P\rightarrow E$ the verticality condition is relaxed. The gauge transformations in $P\rightarrow E$ are allowed to translate within the base space $E$. To see this we write the gauge transformation as $f=L_g$ for some $g\in G$ and $h\in H$,
\begin{equation}
 f(\sigma_{EP}(p,\xi))=\sigma_{EP}(p,\xi')h.
\end{equation}
We consider the gauge transformations of a vector $X$ tangent to a curve passing through the point $u=\sigma_{EP}(p,\xi)$ to find an equation analogous to \Eref{eq:bundlegaugetrans},
\begin{equation}
 f_* X= Q_*(R_{h*}X + (h^{-1}\exd h)^{\#}(X)).\label{eq:compbundlegaugetrans}
\end{equation}
There are two differences between \Eref{eq:bundlegaugetrans} and \Eref{eq:compbundlegaugetrans}. First although we used an element $g\in G$ as our left action, only the component $h\in H$ survives in the expression for the transformation. Second there is an additional pullback by $Q=R_h^{-1}\circ L_g$ present in the transformation. These differences will conspire to ensure the coframe transforms as a contravariant Lorentz vector. To see this we first need to introduce the connections. There appear to be two bundles on which we can put a connection, i.e.\ $P\xrightarrow{\pi_{PE}} E$ and $E\xrightarrow{\pi_{EM}} M$. But since the overall bundle is $P\rightarrow M$, there is really only one connection 1-form $\omega$. The two connection 1-forms, $P\xrightarrow{\pi_{PE}} E$ and $E\xrightarrow{\pi_{EM}} M$, are the ''shadow`` of this connection pulled back to their respective base spaces,
\begin{equation}
 \mathcal{A}_M=\sigma_{MP}^*\omega,\hspace{1cm}\mathcal{A}_E=\sigma_{EP}^*\omega.
\end{equation}
The connection on $E$ can be further pulled back to the base space $M$ so that the two ``shadow" connection 1-forms satisfy~\cite{Tresguerres2002},
\begin{equation}
 A_{M}=\sigma_{ME}^*A_E=\sigma_{ME}^*\sigma_{EP}^*\omega.
\end{equation}
Over the bundle sector $P\rightarrow E$ we can split the total connection into a sum of two components $\omega=\omega_{R}+\omega_{T}$, the subscript $R$ denotes Lorentz rotations and the subscript $T$ denotes translations. Analogous to \Eref{eq:generalconnection} we have,
\begin{equation}
 \omega_{R}=h^{-1}(\exd +\pi^*_{PE}\mathcal{A}_{R}) h,
\end{equation}
where $A_{R}$ is the local form of the connection of the Lorentz connection on the base space $E$. Additionally we have,
\begin{equation}
\omega_{T}=h^{-1}\pi^*_{PE}\mathcal{A}_{T} h.
\end{equation}
Where $\mathcal{A}_T$ denotes the local form of the translational connection on $E$. Applying the total connection ($\omega=\omega_{R}+\omega_{T}$) to \Eref{eq:compbundlegaugetrans}, pulling the result to the base space $E$ and equating terms based on their expansions in the Lie algebras of the translations and Lorentz rotations we arrive at the gauge transformation of the gauge fields,
\begin{equation}
 h^{-1}\exd h + h^{-1}\mathcal{A}_{R}h=\mathcal{A'}_{R}, \hspace{.2cm} h^{-1}\mathcal{A}_{T}h=\mathcal{A'}_{T}.
\end{equation}
The infinitesimal variation can be computed from $\delta A= A - A'$ where $A'$ is the gauge transformed form of $A$. We expand the group transformations $h=e^{i\Lambda_{\alpha\beta}\epsilon^{\alpha\beta}}\approx I+ i\Lambda_{\alpha\beta}\epsilon^{\alpha\beta}$ to arrive at,
\begin{equation}
 \delta \mathcal{A}_{T}^{i}=-\mathcal{A}_{T}^{j}\tensor{\epsilon}{^{i}_{j}}.
\end{equation}
This is exactly the infinitesimal variation of a Lorentz transformation. Breaking the bundle $P\rightarrow M$ to $P\rightarrow E\rightarrow M$ gives the right transformation properties of $\mathcal{A}_{T}$ while leaving it to still be identified as a gauge potential~\cite{Tresguerres2002}. This gauge potential ($\mathcal{A}_{T}$) when pulled back to the base by a canonical local trivialization $\sigma_{ME}(p)=\phi(p,e)$ is identified as $\sigma_{ME}^*\mathcal{A}_{T}=\hat{\theta}$, i.e.\ the coframe. We can decompose the coframe as $\hat{\theta}^{i}=\tensor{e}{^{i}_{\mu}}\exd x^{\mu}$ and so can write the metric as,
\begin{equation}
 g_{\mu\nu}=\tensor{e}{^i_{\mu}}\tensor{e}{^j_{\nu}}\eta_{ij}.
\end{equation}
A further indication that composite gauge theories describe gravitation comes from expanding the tetrad in terms of the spacetime connection~\cite{Tresguerres2002},
\begin{equation}
 \tensor{e}{_{\mu}^{i}}=\partial_{\mu}\xi^{i}+\tensor{\mathcal{A}}{_{R}_{\mu}_{j}^{i}}\xi^{j}+\tensor{\mathcal{A}}{_{T}_{\mu}^i}.
\end{equation}
We can see that the tetrad has internal structure in terms of the dynamic gauge fields as expected since the Christoffel connection was defined in terms of the metric.\\

\textit{Bundles necessary to describe gravity ---}To determine whether composite bundles provide a consistent formulation of gravitation, we should stipulate what sort of objects we expect the construction to reproduce.  This point has been slighted in the literature. The requirements arise from considering what bundles are relevant to general relativity. To formulate general relativity via a gauge principal we must show how the tangent bundle and the frame bundle arise as a consequence of local symmetry. The tangent bundle is a vector bundle over a manifold $M$ denoted $TM\equiv E(M,\reals^n,Gl(n,\reals))$. Its structure group is the general linear group and its fiber space is an $n$ dimensional Euclidean space. Associated with the tangent bundle is the frame bundle, a principal fiber bundle denoted $\mathcal{F}M\equiv P(M,Gl(n,\reals))$. The tangent bundle is the associated vector bundle to the frame bundle. A Lorentzian metric on a space $M$ is an inner product operation on the tangent bundle $\eta:TM\otimes TM \rightarrow \reals$. Introducing a Lorentzian metric on the tangent bundle reduces the structure group of the tangent bundle and the frame bundle from the general linear group to the orthogonal group, $Gl(n,\reals)\rightarrow O(1,n-1)$ (supposing that there is one timelike dimension and $n-1$ spatial dimensions). So we are left with $TM=E(M,\reals^n,O(1,n-1))$ and $\mathcal{F}M=P(M,O(1,n-1))$. We can reduce the symmetry group and further extend it as we continue to remove topological obstructions. But it is these two bundles which a gauge theory of gravitation must reproduce. The removal of topological obstructions will be the topic of the following section. However we are still left with the question, how do the frame bundle and its associated vector bundle arise in the context of composite gauge theories of gravitation? \\

\subsection{The general topology of composite bundles}
To answer this question we need to go to back to how and when can we split a principal bundle into a ``tower'' of bundles. The basic idea, as Tresguerres applies it, comes from proposition 5.5 and 5.6 of Kobayashi and Nomizu. 
\begin{mdframed}
    \textbf{Proposition 5.6:} The structure group $G$ of $P(M,G)$ is reducible to a closed subgroup $H$ if and only if the associated bundle $E(M,G/H,G,P)$ admits a cross section $\sigma:M\rightarrow E=P/H$~\cite{Kobayashi}. 
\end{mdframed}
And we are able to identify $E=P(M,G)/H$ due to proposition 5.5 which states, 
\begin{mdframed}
    \textbf{Proposition 5.5:} The bundle $E=P\times_GG/H$ associated with P with standard fiber $G/H$ can be identified with $P/H$ as follows. An element of E represented by $(u,a\xi_0)\in P\times G/H$ is mapped into the element of $P/H$ represented by $ua\in P$ where $a\in G$ and $\xi_0$ is the origin of $G/H$, i.e, the coset $H$~\cite{Kobayashi}.
\end{mdframed}
This proposition assures us that the associated bundle $E$ can be seen to be exactly the total bundle $P$ quotient-ed by the closed subgroup $H$. Reducible in these propositions has a very specific meaning, provided there exists an embedding $f:P'(M,G')\rightarrow P(M,G)$ then the image $f(P)$ is a subbundle and we say $G$ is reducible to $G'$~\cite{Kobayashi}. An understanding of what reduced means is crucial for the identification of the topological structures of classical spacetime. The global section of $E$ leads to a subbundle of $P(M,G)$ given by,
\begin{equation}
 Q(M,H)\equiv \{u\in P(M,G) | \pi_{PE}(u)=\sigma(\pi(u))\},\label{eq:Framedef}
\end{equation}
for $\pi:P\rightarrow M$. The proof of the existence of the subbundle $Q(M,H)$ is provided in Kobayashi and Nomizu in the proof of proposition 5.6~\cite{Kobayashi}. 
There is a further bundle we will need. For every principal bundle there is an associated vector bundle. So we can also construct $Q'=Q\times_{H} \reals^n$. \\

\textit{Identification of the composite bundles and the tangent and frame bundles ---} As a result of requiring a global section to split $P(M,G)$ we have a collection of bundles to work with which are displayed in \tref{tab:lotsobundles}. 
\begin{table}[ht]
\caption{\label{tab:lotsobundles} The collection of bundles formed during a composite bundle construction and their projections.}
\begin{indented}
\lineup
\item[]\begin{tabular}{cc}
\br
 Bundle & Projection \cr
 \mr
 $P(M,G)$ & $\pi$ \\
 $P(E,H)$ & $\pi_{PE}$ \\
 $E(M,G/H,G,P)$ &$ \pi_{EM}$ \\
 $Q(M,H)$ & $\pi_q$ \\
 $Q(M,H)\times_H \reals^n$ & $\pi_{qA}$ \\
 \hline 
 \end{tabular}
 \end{indented}
\end{table}
This collection of bundles is our primary result, here we find the appearance of the frame and tangent bundle in composite gauge theory. For $G=ISO(1,3)$ the Poincar\'e group, $H=O(1,3)$ the Lorentz group and $G/H=\reals^4$ the bundle $Q$ is diffeomorphic to the frame bundle $\mathcal{F}M\cong Q(M,O(1,3))$. The associated vector bundle to $Q$ is diffeomorphic to the tangent bundle $Q'=Q\times_{O(1,3)}\reals^4\cong TM $. The global sections of the bundle $E$, which assure us a subbundle $Q\subset P$, also act to connect the translational and rotational gauge degrees of freedom of $P(E,H)$ and $E\cong P/H$ to the spacetime bundles $\mathcal{F}M$ and $TM$. Furthermore, there is a $1:1$ correspondence between sections of $E$ and the subbundles $Q(M,H)$. If we did not require a global section of $E\cong P/H$ then we would have no way of connecting the gauge bundle with the frame bundle meaning that the gauge bundle would not influence spacetime. This point seems to be missed in the literature.

\section{Topological Obstructions}\label{chap:classes}
The construction of a consistent composite bundle requires that the associated bundle $E$ admit a global section $\sigma$. Typically a global section of a bundle implies triviality of the bundle, hence it may be written directly as a Cartesian product. Meanwhile the global structure of the bundle in general may be characterized by cohomology groups. These groups provide the topological information about the bundle which can obstruct the creation of global features of a fiber bundle. As such cohomology is a powerful language to speak about global properties of the manifolds we are interested in using to describe our physical reality. Of particular interest to us in this work is~\cite{Flagga2002,Flagga2004} who provided further details about the \v{C}ech cohomology and the associated Stiefel-Whitney classes. This cohomology is built entirely from the transition functions of a bundle. Since a bundle is trivial iff we can choose every transition function to be trivial it is natural to suspect a cohomology built from the transition functions to be an appropriate tool to characterize the bundles global structure. In this section we will discuss the global structure of the composite bundle approach to Poincar\'e gauge theory in two aspects, first we will address the conditions under which we can construct a composite bundle to begin with, we will then continue onward providing a continuation of~\cite{Flagga2002,Flagga2004}, pointing out that if we impose basic requirements on our bundle construction then we are lead to impose conditions on our manifold which arise in consistent descriptions of theories of extended objects. \\

\textit{Sufficient condition for the composite bundle construction ---} Let's begin with the first point. In the previous section we ended with the conclusion that we must have a global section of $E\cong P/H$. However, we didn't answer if this was possible to begin with. That is, over which manifolds can we find global sections of $E$? It turns out that there are conditions which guarantee its existence. Theorem 5.7 of Kobayashi and Nomizu gives that for $E(M,F,G,P)$ if the base space $M$ is paracompact~\footnote{A topological space $M$ is paracompact if every open covering of $M$ is such that each point of $M$ is covered by a finite number of sets in the cover~\cite{nakahara}.} and the fiber space, here $F=G/H$ is diffeomorphic to an Euclidean space any section defined over a closed subset of $M$ can be extended to the entire space $M$~\cite{Kobayashi}. Therefore, we are led to the idea that, if we wish to take seriously the composite bundle construction of general relativity, we are guaranteed its existence if the base manifold is paracompact. Notice, this is not a necessary condition. However this condition is sufficient to provide us with composite bundle construction. We will therefore take our base manifold to be paracompact~\footnote{A compact Lorentzian manifold contains closed time-like curves~\cite{Hawking:1973uf}.}. \\

\textit{Characteristic classes ---} Interestingly, this sufficient condition leads us directly into the second aspect of our discussion in this section. A primary theme of~\cite{Flagga2002,Flagga2004} was to place conditions on the global structure of spacetime. Our local spacetime geometry is endowed with a pseudo-Riemannian, or Lorentzian metric. A manifold admits a Lorentzian metric iff it is paracompact and admits an everywhere non-vanishing continuous, up to a sign, direction field~\cite{Visser:1995cc}. It is highly interesting that the composite bundle construction leads us to consider, again as a sufficient condition, a paracompact manifold, preparing us for a Lorentzian signature metric. 

What if we want to ensure that our manifold may have a Lorentzian signature? Then we must have the second of these conditions also satisfied, hence we must be able to define such a direction field, ensuring that we can everywhere label past and future directed vectors, implying our spacetime is time-orientable. Likewise, it is further pointed out in~\cite{Flagga2002,Flagga2004} that the standard model of particle physics is not consistent with spacetimes which are not space-orientable. That is, there exists an additional, everywhere non-vanishing direction field $d_{spatial}$ and three form $\omega_{spatial}$ such that $\omega_{spatial}(d_{spatial})=0$.  If we were to assume this was also true then the consequence of these arguments and assumptions is that the structure group of our bundle $O(1,3)$ is further reduced from $O(1,3)$ to $SO(1,3)^+=\{A\in O(1,3)|\det A=1, A^0_0>0\}$, the proper Lorentz group.

This set of conditions concerning the orientability of the base spacetime manifold is neatly captured in terms of the objects known as Stiefel-Whitney classes $w_r\in H^r (M; \mathbb{Z}_2)$, which are elements of the \v{C}ech cohomology (in the appendix~\ref{app:cech} we compile some basic information about the \v{C}ech cohomology). The non-triviality of these classes represent obstructions to the creation of certain global structures on bundles. Indeed, as we just mentioned, and will discuss below in more detail, the first of these classes is associated with the orientability of the space~\cite{Milnor,Husemoller,nakahara}. The first two of these classes have long had interpretations, while the third and fourth have only, within the last two decades, been given a physical interpretation. Before discussing how this applies to the composite bundles we are considering, we first will recall some more information about these characteristic classes. To that end we remind the reader that if $M$ is a manifold and $TM\xrightarrow{\pi}M$ is its tangent bundle, the Stiefel-Whitney classes of $M$ are the same as the those of its tangent bundle~\footnote{Proof of this statement can be found in Milnor and Stasheff~\cite{Milnor}}. \\

\textit{The first Stiefel-Whitney class ---} Let $TM$ have a Riemannian structure provided by $g:\pi^{-1}(p)\otimes\pi^{-1}(p)\rightarrow \reals$. Recall that $\pi^{-1}(p)\cong F$, i.e.\ the Riemannian structure on $TM$ is then a mapping $g:TM\otimes TM:\rightarrow\reals$. In general relativity we could choose a local frame as a rotation of the natural basis in the tangent space. If we consider the set of all of these choices we can build a principal fiber bundle called the frame bundle. The frame bundle has the tangent bundle as an associated bundle and is written $\mathcal{FM}(M,Gl(n,\reals))$. At first we have a choice of $\tensor{e}{^{\mu}_{i}}\in Gl(4,\reals)$. However we require that the frame $\hat{e}$ be orthonormal with respect to the metric. This requirement reduces our set of choices from a $Gl(n,\reals)$ rotation to an $O(n)$ (or $O(1,3)$ for Lorentzian spacetimes) rotation. The reduction in structure group means on overlapping open neighborhoods $U_i\cap U_j$ that a choice of frame over $U_i$ is related to the choice in frame of $U_j$ by $e_{\alpha}=t_{ij}e_{\beta}$ with $t_{ij}\in O(n)$ as the transition function. Let $f$ be the determinant function, we have $f(t_{ij})=\pm1$, i.e.\ $f$ is indeed valued in $\integers_2$.
On a triple intersection $U_i\cap U_j\cap U_k$ the transition functions ($t_{ij}$,$t_{jk}$,$t_{ki}$) are required to satisfy the cocycle condition $t_{ij}t_{jk}t_{ki}=I$~\cite{nakahara,frankel}. Using the cocycle condition we can act the boundary operator on our cochain function to find $\delta f(i,j)=1$, so $f\in Z^1(M;\integers_2)$ and defines an equivalence class $[f]\in H^1(M;\integers_2)$. This first class $w_1=[f]$ is the first Stiefel-Whitney class. It follows that $M$ is orientable iff $w_1$ is trivial, see Nakahara~\cite{nakahara} for proof. See also~\cite{Flagga2002} for a list of equivalent conditions for the triviality $w_1$.\\

\textit{The second Stiefel-Whitney class ---}  Defining spinors on a space requires a lifting to  the covering group of $SO(1,3)^+$ for 4-dimensional Lorentzian spacetimes denoted~\footnote{In Euclidean signature we could also consider $SO(n)$ for which the covering group is denoted $PIN(n)$.} $SPIN(n)$ which is the group $SL(2,\mathbb{C})$. The second Stiefel-Whitney class describes the obstruction to defining this lifting. Since we can reconstruct a bundle from its transition functions, consider a manifold $M$ whose tangent bundle is orientable i.e.\ the first Stiefel-Whitney class is trivial, and let $\{t_{ij}\}$ be the set of transition functions of the associated principal frame bundle $\mathcal{F}M$. If we let $\psi:SL(2,\mathbb{C})\rightarrow SO(1,3)^+$ be the typical double covering of the group $SO(1,3)^+$ we can define a complementary set of transition functions $\{\tilde{t}_{ij}\}$ such that $\psi(\tilde{t}_{ij})=t_{ij}$ and,
\begin{equation}
 \tilde{t}_{ij}\tilde{t}_{jk}\tilde{t}_{ki}=I,
\end{equation}
on a triple intersection $U_i\cap U_j\cap U_k$. The set of $\{\tilde{t}_{ij}\}$ defines a spin bundle over $M$ denoted $D(M,\mathbb{C}^4,SL(2,\mathbb{C})\oplus \overline{SL(2,\mathbb{C})},\pi_D)$ and $M$ is said to admit a spin structure. Consequently we have that a manifold $M$ admits a spin structure iff the second Stiefel-Whitney class $w_2\in H^2(M,\mathbb{Z}_2)$ is trivial~\cite{nakahara}, this is often referred to by saying that $M$ is parallelizable. Importantly, an equivalent condition for the parallelizability of a manifold is that $M$ possesses a global section in the frame bundle $FM$ and hence, by extension, a global section of its tangent bundle $TM$.\\

\textit{The third Stiefel-Whitney class ---} Having established a spin structure,  in curved space, as in flat space, we locally define a chirality operator $\gamma_5$. In curved space this takes the form~\cite{Flagga2002}
\begin{equation}
    \gamma_5=\frac{\omega_{[\mu\nu\lambda\kappa]}}{4!}e^{\mu}_ae^{\nu}_be^{\lambda}_ce^{\kappa}_d\gamma^a\gamma^b\gamma^c\gamma^d
\end{equation}
with $\gamma^a$ the flat space $\gamma$ matrices and $\omega_{[\mu\nu\lambda\kappa]}$ is a four-form. We can go further, following~\cite{Flagga2002} and define projection operators onto left and right handed spinors which act on Dirac spinors $\psi$ (sections of the spin bundle $D(M,\mathbb{C}^4,SL(2,\mathbb{C})\oplus \overline{SL(2,\mathbb{C})},\pi_D)$) to give $\psi_L$ and $\psi_R$. These left and right handed spinors become sections of the left and right hand Weyl bundles $W(M,\mathbb{C}^4,SL(2,\mathbb{C}),\pi_W)$ and $\overline{W}(M,\mathbb{C}^4,\overline{SL(2,\mathbb{C})},\pi_{\bar{w}})$. The result is that spin bundle may be decomposed as a Whitney-sum bundle $D=W\oplus \overline{W}$. The requirement that we can split this spin bundle as a Whitney-sum globally is equivalent to the ability to define the four form $\omega$ throughout the whole manifold. And furthermore the authors of~\cite{Flagga2002} show that this is equivalent to the requirement that, $w_3\in H^3(M,\mathbb{Z}_2)$ is trivial. Hence in four dimensional spacetime chirality is globally defined iff $w_3\in H^3(M,\mathbb{Z}_2)$ is trivial. A note, $w_3$ is trivial iff $w_1$ and $w_2$ are trivial.  \\

\textit{The fourth Stiefel-Whitney class ---} In four dimensions the fourth Steifel-Whitney class is the top class, i.e. $w_r$ for $r>4$ all vanish. This introduces some slight complications, since there are different interpretations of the top even class when the manifold is even or odd. In our case~\cite{bredon1993topology}, with an even manifold we can view the frame bundle as 4-plane bundle and consequently $w_4$ is the mod 2 reduction of the Euler class $e(M)\in H^4(M,\mathbb{Z})$. Both class are trivial if the manifold possesses a globally defined nowhere-zero section. Furthermore, this section can be found to be related to our original motivation for studying the global structure of $P(M,G)$, for if $M$ is stably causal, then $e(M)$ and $w_4$ are trivial and the Euler characteristic $\chi(M)=0$. Notice that if $M$ is stably causal then there exists a global $\mathbb{R}$-valued function such that its gradient is everywhere time-like. Hence the gradient, which is given from the components of the 1-form $(df)_\mu \mathrm{d} x^\mu=g_{\mu\nu}V^\nu\mathrm{d} x^\nu$, satisfies $V^2<0$ for all points in $M$. That is, it is everywhere non-zero an provides a global section in $TM$.  \\

For our four dimensional geometry we have only the four Stiefel-Whitney classes, each gives information about global structures that can be present. In Table.~\ref{tab:SWC} we display a summary of the interpretations of each Stiefel-Whitney class).
\begin{table}[ht]
\caption{\label{tab:SWC}A list of the Stiefel-Whitney classes and their interpretations as obstructions to topological entities over a manifold $M$.}
\begin{indented}
\item[]\begin{tabular}{cc}
\br
Stiefel-Whitney class & What is it obstructing? \cr
\mr
 $w_1$  &  Orientability \\
 $w_2$  &  Spin\\
 $w_3$  &  Chiral Spinors\\
 $w_4$  &  Causality \\
 \hline
\end{tabular}
\end{indented}
\end{table}
Naturally, not all manifolds are such that all of them vanish or are trivial. For instance while a vanishing $w_2$ implies a vanishing $w_1$ and hence $w_3$ is also vanishing, a vanishing $w_1$ does not imply a vanishing $w_2$, complex projective spaces being an excellent example.  In the years since~\cite{Flagga2002,Flagga2004} there has been many developments in the study of characteristic classes, especially in relation to string theory~\cite{Stolz_Teichner_2004,Waldorf2009,Sati2009,Sati:2010ss,Sati:2010dc,waldorf2011loop,Sati:2014yxa,Sati:2018qrz,Fiorenza:2020hiq}. In particular these works further developed old results~\cite{Killingback1987,Witten:1988hc,mclaughlin1992orientation} emerging from anomaly cancellation in two-dimensional supersymmetric worldsheet theories underlying sypersymmetric string theory. This led to the creation of multiple new interpretations of characteristic classes as obstructions and the definition of string~\cite{Stolz_Teichner_2004,Waldorf2009}, fivebrane~\cite{Sati2009} and ninebrane~\cite{Sati:2014yxa} structures and twisted variants. Although it appears on the surface that these would not be relevant here, this is not necessarily true and hence we include a brief discussion of the string structure since it will make an appearance here. \\

\textit{String Structures ---} From a purely topological point of view a string structure is an additional structure on a principal spin bundle $P$, it is a cohomology class $\xi\in H^3(P,\mathbb{Z})$ (see~\cite{Waldorf2009}). Alternatively, this can be defined as a lift of the structure group from $Spin(n)$ to a three connected extension known as the string group~\cite{Stolz_Teichner_2004} (which can be viewed as an infinite or finite Lie 2-group, see ~\cite{hha/1201127333,schommer2011central} for more details). Importantly, we have from~\cite{Stolz_Teichner_2004,Waldorf2009} that $P$ admits a string structure if and only if the first fractional Pontryagin class $\frac{1}{2}p_1(P)\in H^4(M,\mathbb{Z})$ vanishes. This means that not only must $p_1$ be trivial but the cohomology group must be torsion free~\cite{Sati2009}. The relevance of lifting to the string group in physics was pointed out by Killingback~\cite{Killingback1987} who displayed the role the of the 2-form $B$ and its 3-form field strength $H$ in the cancellation of spacetime anomalies. The 3-form obeys,
\begin{equation}
 \exd H=Tr(F^2)-Tr(R^2) \label{eq:consistent}
\end{equation}
for $F$ a Yang-Mills field strength and $R$ the curvature 2-form. Integrating this over a closed subset of base space $M$ results in a consistency condition that the two Pontryagin classes of the tangent bundle of spacetime and of either $SO(32)$ or $E_8\times E_8$, be equivalent. It is important here to point out that this involves the gauge bundle on the target space of the string model and necessarily involves the field strengths of the gauge degrees of freedom $F$ and the field strength of the tangent bundle $R$. In what follows below we will be concerned with string structure of the manifold only and hence imposes the condition that $\frac{1}{2}p_1(TM)=0$. 
\\

\textit{Application to the construction of the composite bundle ---} Now that we have collected the characteristic classes needed we can discuss how this applies to composite bundles. Of interest to us is a paracompact base manifold. Unfortunately, cohomology becomes more complicated when the manifold $M$ is non-compact. To work with a sensible cohomology theory on non-compact spacetimes we need cohomology with compact support (see for instance~\cite{Bott82} for more information), in essence we will have cochains that vanish outside of a compact set. Although this is a complication, we will still be able to define all of the ordinary objects of interest. In order to not distract from the main discussion in~\ref{app:compact_support} we provide the necessary information to justify the remaining conclusions of this section. In what follows we will discuss the cohomology as we normally would, although the reader should keep in mind we are actually discussing compactly supported cohomology and refer to the information contained in~\ref{app:compact_support}. 

Let us recall that we found that a sufficient condition for $E\cong P/H \cong E(M,G/H,G,P)$ to exist is that $M$ be paracompact and $G/H\cong R^n$ with $G=ISO(1,3)$ and $H=SO(1,3)$. If this is true then there exists a section $\sigma: M\rightarrow E$ and hence $E\cong M\times_{G} G/H$ or $M\times_{ISO(1,3)} \mathbb{R}^4$. Since this section exists the bundle is trivial, hence all transition functions are trivial, and for the sake of argument we can consider a single local trivialization under which $\sigma$ is an assignment of points $X\in \mathbb{R}^n$ for each point in the base manifold. Since the fiber space is $\mathbb{R}^n$, the assignment must be non-zero for every $p\in M$. Hence we have found we can construct a global section of a vector bundle over $M$. However this section is precisely what is needed to construct a global section of $TM \cong Q\times \mathbb{R}^n$ which in the notation above would be $TM(M,\mathbb{R}^4,H,Q)$. For each choice of $\sigma$ there is a 1-1 correspondence with reduced bundles $Q\subset P$ and this set of possible bundles is precisely all the ways in which we can transform an assignment of $X$ by translation. Therefore we can map $\iota:TM(M,\mathbb{R}^4,H,Q) \rightarrow E(M,G/H,G,P)$ by the inclusion map and there exists a section of $q_{TM}:M\rightarrow TM$ such that $\iota(q_{TM}(x)) =\sigma(x)=(x,X)$ for a choice of $\sigma$. 

We then arrive at the conclusion that the sufficient condition for $E$ to exist gives a requirement that the tangent bundle have a global section. The implication then, is that the Euler class is trivial. With integer coefficients the Euler class is carried by the natural homomorphism to the top Stiefel-Whitney class and so it too is trivial~\cite{Milnor}. If instead we work with real coefficients in the de Rham cohomology the Euler class of a even dimensional manifold squares to the first Pontryagin class $p_1$, requiring it to trivialize. Notice, that for this to be the true, it must be the case that the group action of the tangent space, and hence the fiber space of the frame bundle, is reduced, from $O(1,3)\rightarrow SO(1,3)^+$. Note we can do this since we are free to choose a section whose motion in the fiber space consists of orbits of $SO(1,3)^+$. That is, the bundle must be orientable, and since the tangent space carries an every non-vanishing section the frame bundle must be trivial i.e. the first two Stiefel-Whitney classes are trivial, and hence so is the third. It appears that this demand of breaking the Poincar\'e bundle into its substructure places severe topological constraints on the spacetime manifold.

Before closing this section we note that a cohomology group $H^r$ can be decomposed into two pieces, a free piece and a torsion piece. The vanishing of the first fractional Pontryagin class is the condition that $H^4(M;\integers)$ be decomposed as only a free piece and the characteristic class $p_1\in H^4(M;\reals)$ be trivial. We already argued that the first Pontryagin class is trivial. What is left to show is that the decomposition of the cohomology group has only a free piece. The idea here is that a finitely generated abelian group is the direct sum of a free abelian group of finite rank and a finite abelian group~\cite{Hatcher2002}. A free group has a basis in which we can represent an element in terms of the basis elements and for a group to be finitely generated there must be a finite number of basis elements. While the finite abelian subgroup is the torsion subgroup. Loosely, the torsion subgroup is made up of all elements with finite order. 

Provided that $M$ is a connected and compact manifold, the homology groups are finitely generated. And provided there is an orientation (which there is), we can use a theorem from Hatcher~\cite{Hatcher2002}, if $M$ is a compact connected manifold of dimension $n$ then $H_{n-1}(M,\integers)$ is trivial if $M$ is orientable. In this case using the universal coefficient theorem and Poincar\'e duality we have $H^4(M,\integers)=H_4(M,\integers)\oplus H_3(M,\integers)$. However $H_3(M,\integers)$ is trivial so we have exact Poincar\'e duality, i.e.\ $H^4(M,\integers)=H_4(M,\integers)$. Along with the previous result of the triviality of the integral cohomology class we conclude that the manifold $M$ admits a lifting to $\text{STRING}(n)$. Notice, every compact manifold is paracompact and hence we can still define the composite bundle structure.

Unfortunately, as mentioned earlier, compactness is too strong a condition and leads to closed timelike curves in Lorentzian signature manifolds. Hence we must argue this to hold for the paracompact case. The necessary aspect of Hatcher's lemma~\cite{Hatcher2002} is that the homology groups are finitely generated. Thus to use the same reasoning we must have that the compactly supported homology groups we have created are finitely generated. Fortunately we have that provided that a space $M$ has a finite good cover (if all finite intersections of sets in the cover are diffeomorphic to $\mathbb{R}^n$) then $dim H^r_c(M;\reals)<\infty$ \cite{Bott82}. Fortunately, since $M$ is a smooth paracompact manifold it admits a finite good cover. 

With this result we can now use the Poincar\'e duality for compactly supported cohomology to say that the spaces $H^r_c$ are isomorphic to $H_{n-r}$ which is to say the dimension of the homology groups are finite for these spaces~\cite{Hatcher2002,Husemoller}. Using lemma 3.27 from Hatcher, $H_i(M,M-A;\reals)=0$ for $i>n$ on a compact subset $A$ of an $n$ dimensional manifold $M$, we see that there are finitely many homology groups~\cite{Hatcher2002}. With a finite number and finite dimension, our homology groups are finitely generated. Then using the universal coefficient theorem $H_r(M;\reals)\cong H_r(M;\integers)\otimes \reals$ and so the homology groups with integer coefficients are finitely generated. We can finally use the flavor of corollary 3.28 of Hatcher~\cite{Hatcher2002} to say that $H_{r-1}(M;\integers)$ has trivial torsion subgroup. The result of all of this formalism is that in the realistic case of a paracompact base space we also have the trivialization of first Pontryagin class and the fourth cohomology class has no torsion. 

Notice, here we have two results, the first, is that if a four dimensional manifold, or Lorentzian signature, is in agreement with the conditions put forth by the authors of~\cite{Flagga2002,Flagga2004} on a reasonable spacetime topology, that is all four Stiefel-Whitney classes are trivial and the Euler class is trivial, then so too is the first Pontryagin class and the fourth cohomology class has no torsion, hence the manifold is forced to admit a string structure. Likewise, if M is an oriented, four dimensional manifold of Lorentzian signature that is stably casual, the fourth Stiefel-Whitney class is trivial and hence so is the second and third and furthermore the manifold admits a string structure. Second, a sufficient condition for the construction of a composite bundle formulation of gravity in four spacetime dimensions is we are forced to have all of the Stiefel-Whitney classes be trivial, as well as the Euler class and Pontryagin be trivial and hence the manifold admit a string structure.

\section{Conclusions}
We have found that the composite gauge theory formulation of general relativity requires that the Stiefel-Whitney, Euler and first rational Pontryagin classes of spacetime be trivial. This set of topological restrictions is consistent with the conclusion that spacetime must be a string manifold, i.e.\ admit a string structure. This comes somewhat as a surprise since this analysis has been purely classical and the usual motivation for extended degrees of freedom like strings (and hence the admittance of string structures for spacetime) in gravitational theories is to achieve quantum consistency. Perhaps this result might have been anticipated since every viable gauge theory of gravity seems to necessitate the presence of torsion in a manner reminiscent of the inevitable inclusion of the Kalb-Ramond field (whose field strength mimics torsion) in the massless multiplet that includes the graviton in string theory.

It should be stressed that we did not start out to investigate gauge theories of extended objects as in the program of so called higher gauge theory ~\cite{higherTPG}. Instead our starting point was the minimal bundle formulation accommodating both the Lorentz and translational symmetries underlying Poincar\'e gauge theory. However given the resulting connection to extended degrees of freedom, it would be an interesting task to connect the composite bundle framework with developments in higher gauge theory, see for instance the recent work on the categorification of string structures of principal bundles~\cite{Berwick-Evans:2021hqb}.

It is also interesting to pause and distinguish how this composite construction differs from other physical constructions of gravity. We followed closely~\cite{Flagga2002,Flagga2004} whose goal was not to understand gauge theories of gravity. Rather, their goal was to provide physical interpretations of topological constraints on a spacetime, providing requirements for a physically reasonable spacetime. One obvious conclusion of our work is a continuation of theirs. We can briefly forget the discussion of the composite bundles and focus soley on the topology of spacetime. If we require a spacetime $M$ to be a four dimensional and archwise connected smooth manifold with Lorentzian signature with volume form (and hence spacetime, space and time orientable) on which spinors may be defined and is stably causal then all four Stiefel-Whitney classes vanish. Furthermore the Euler class vanishes, and the as discussed here this implies that the first fractional Pontyagin class is also trivial. Hence in general, we have shown how these considerations of physically reasonable spacetimes already imply that our physical spacetime manifold admit a string structure. What we have shown in addition to this, is that if we consider writing a four dimensional general relativistic theory as a composite bundle structure of Poincar\'e gauge theory that a sufficient condition for its existence is that all of the characteristic classes of our manifold are trivial and hence the spacetime manifold admit a string structure.

There are many open avenues for further exploration. Our analysis has largely focused on the simplest realization of general relativity and hence requires no more than four spacetime dimensions. In string theory, we not only encounter topological restrictions based on string structure, but also restrictions based on the admittance of a five-brane structure~\cite{Sati2009}. Clearly the five-brane structure is not relevant for four dimensional settings, but investigation of the consequences of composite bundle consistency in higher dimensions could very well lead to the requirement of five-brane structure as well. 

Yet another consideration is the inclusion of internal gauge symmetries in the composite formulation. The consistent splitting of the total bundle into a composite structure could likely lead to related conditions between topological invariants of the base spacetime and of the internal gauge bundle. These results might very well tie in with the spacetime and gauge anomaly cancellation mechanisms underlying the consistency of Type I and heterotic strings theories just as in \Eref{eq:consistent}.    

As a final note, a spin structure is the relevant for the proper definiton of spinor wavefunctions on a spacetime manifold as a section in a vector bundle with a particular representation. This representation does not lead to a representation of $SO(d)$ but to the covering group and hence we need to lift of the structure group to $Spin(d)$. This is very much the analogus story for strings (for more details see the nice review~\cite{Waldorf:2023zmt}). However, since the energy scale required to resolve stringy physics is still very much beyond our ability it is likley this does not have any immediate consequence for currently observable physics.

\section*{Acknowledgements}
This work was completed while Casey Cartwright was a M.Sc. student at the Colorado School of Mines, Golden, Colorado 80401, USA as part of a Masters thesis under Alex Flournoy's supervision. The work in its current form is dedicated to Alex T. Flournoy, January 19, 1974 --- November 3, 2023, a phenomenal teacher, mentor and friend. This article was prepared for submission while Casey Cartwright was supported by the Netherlands Organisation for Scientific Research (NWO) under the VICI grant VI.C.202.104.

\appendix 
\section{More on fiber bundles}\label{app:fiber}
\begin{figure}[ht]
\begin{center}
 \includegraphics[width=3in]{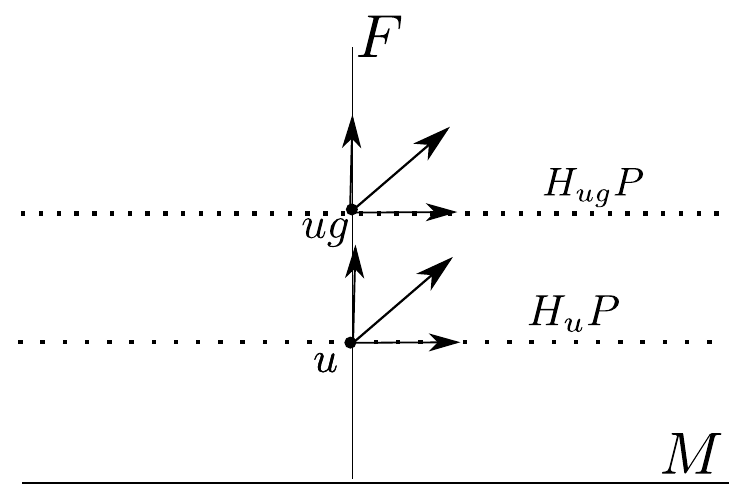}
\end{center}
\caption{\label{fig:horisubs}A vector in the total space $P$ shown decomposed into its vertical and horizontal components. The vector is additionally shown right translated ''up`` through the bundle.}
\end{figure}

In this appendix we collect a few more facts about fiber bundles for completeness in order to contrast with the composite construction.
\textit{Connections and splitting the tangent space ---}Another important consideration is the transport of a vector through the total bundle space. Of concern is if the vector in the bundle is parallel with the base or with the fiber. This can be accomplished by defining a splitting of the bundle tangent space $T_uP$ into vertical and horizontal subspaces $V_uP$ and $H_uP$ respectively (\fref{fig:horisubs}). 
The requirements of the split are, 
\begin{enumerate} 
\item{ $T_{u}P =  H_{u}P\oplus V_{u}P$.}
\item{A smooth vector field $X$ in $P$ can be written as 
  $X=X^{H}+X^{V}$ for $X^{H} \in H_{u}P$ and $X^{V} \in V_{u}P$. }
\item{$R^{*}_{g}H_{u}P = H_{ug}P$ where $R^*_g$ is differential map induced by the right action of $g$.}
 \end{enumerate}
The first two conditions specify the form of the split and the third condition comes from the idea that these fibers extend over the manifold. Just as any section is a right translation of the canonical local trivialization, horizontal subspaces are related by a right translation. Splitting the tangent space in this way specifies a connection on $P$.

In order to extract the more familiar notion of connection in gauge field theory we define the fundamental vector field, $A^{\#}$, at a point $u\in P$ generated by $A\in \mathfrak{g}$ as,
\begin{equation}
 A^{\#}\mathit{f}(u)= \frac{\exd}{\exd t}\mathit{f}(u e^{At})|_{t=0}, \label{eq:fundamental}
\end{equation}
for any function $\mathit{f}:P\rightarrow \mathbb{R}$. The vector $A^{\#}$ is directed entirely along the fiber. This can be seen by pushing forward a vector $X\in V_uP$ by the projection $\pi$ as $\pi_* X$. With this the connection 1-form $\omega\in \mathfrak{g}\otimes T^*_uP$ is defined so that,
\begin{enumerate}
 \item $\omega(A^{\#})=A$ for $A\in \mathfrak{g}$,
 \item $R^*_g\omega= g^{-1}\omega g$.
\end{enumerate}
The horizontal subspace can then be defined as the kernel of $\omega$,
\begin{equation}
 H_uP=\{X\in T_uP | \omega (X)=0\}.
\end{equation}
Now that we have a connection on the total space we can use a section $\sigma$ to pullback the connection form $\omega$ to the base. The connection form on the base can be written,
\begin{equation}
 \mathcal{A}_i=\sigma_i^*\mathcal{\omega}.
\end{equation}
The connection 1-form $\omega$ is a global quantity defined in the total bundle. We often work in terms of locally measurable quantities, e.g. the strength of the electromagnetic field in a local region of space. It is then more useful to start with a local connection 1-form $\mathcal{A}$ on $U_i\subset M$ on the base and define the connection 1-form $\omega$ as,
\begin{equation}
 \omega= g_i^{-1}\pi^*\mathcal{A}_ig_i +g_i^{-1}\exd g_i. \label{eq:generalconnection}
\end{equation}

\textit{Gauge transformations ---} In the language of bundles a gauge transformation of a principal fiber bundle is a base preserving fiber automorphism~\cite{Tresguerres2002,lord86}. That is $f:G\rightarrow G$ such that $R_g\circ f=f\circ R_g$ and $f\circ \pi = \pi$. The first condition says that $f$ must commute with all right actions of $G$ and so is a left action of $G$, $L_g$~\cite{lord86}. The second condition means that the transformation is directed vertically along the fiber and ensures that the gauge transformation does not induce a diffeomorphism of the base. If we write $f(u)=u \eta(u)$ for $u\in P$ and $g\in G$ and $\eta:G\rightarrow G$ defined as $\eta(u)=u^{-1}gu$ we can determine the effect of a gauge transformation on a vector $X\in T_uP$. Let $\gamma:\mathbb{R}\rightarrow P$ such that $\gamma(0)=u$ and $\dot{\gamma}(0)=X$ (where the dot represents a parameter derivative) we find,
\begin{equation}
 f_* X = \frac{\exd}{\exd t}f(\gamma(t))\bigg|_{t=0} =  R_{g*}X + (\eta^{-1}d\eta)^{\#}(X).  \label{eq:bundlegaugetrans}
\end{equation}
Once we have the action of a gauge transformation on a vector in the total space, we can apply the connection 1-form $\omega$ and pullback the result to the base space. Doing this we obtain the transformation properties of the local connection 1-form $\mathcal{A}$. Using \Eref{eq:bundlegaugetrans} let $\sigma_i:M\rightarrow P$ be a local section over the subset $U_i\subset M$. Applying $\omega$ to \Eref{eq:bundlegaugetrans} we find,
\begin{equation}
 f^*\omega(X) = R^*_g\omega(X) +\omega((\eta^{-1}d\eta)^{\#}(X)) .
\end{equation}
Applying $\sigma_i^*$ to pullback the gauge transformation of the connection 1-form $\omega$ to the gauge transformation of the local connection 1-form $\mathcal{A}$ we have
\begin{equation}
 f^*\mathcal{A}_i=g^{-1}\mathcal{A}_i(X)g+\eta^{-1}d\eta(X).
\end{equation}
For the example of $M=\mathbb{R}^{1,3}$ and $G=U(1)$ we coordinatize the space and expand the local connection in the dual basis as $\mathcal{A}=\mathcal{A}_{\mu}\exd x^{\mu}$. The exterior derivative then takes the familiar form when acting on functions(say $\theta\in \mathscr{F}(M)$), $\exd\theta=\partial_{\mu}\theta\exd x^{\mu}$,
\begin{align}
 f^*\mathcal{A}_i=\mathcal{A'}_i&=\mathcal{A}_{i}+e^{-i\theta}d(e^{i\theta})\nonumber\\ 
  \mathcal{A'}_{i\mu}&=\mathcal{A}_{i\mu}+i\partial_{\mu}\theta.\label{eq:bundleu1}
\end{align}
\Eref{eq:bundleu1} is the familiar gauge transformation of electromagnetism, and so we see that vertical bundle automorphisms are indeed the correct notion of gauge transformations for internal symmetry groups. \\

\section{Review of the \v{C}ech cohomology}\label{app:cech}
To build the \v{C}ech cohomology we follow the same general steps as any cohomology theory and follow closely Nakahara~\cite{nakahara}. The coefficients of the \v{C}ech cohomology are $\mathbb{Z}_2$, which we take to be the multiplicative group of elements $\{-1,1\}$. Analogous to the cocycle group in the de Rham cohomology we define the \v{C}ech r-cochain as a function $f(t_0,\ldots,t_r)\in \mathbb{Z}_2$ defined on the set $U=\cap_{j=0}^{r}U_j$ where the $t_r$ are the transition functions. Additionally we require $f$ be invariant under arbitrary permutation. Let $C^{r}$ denote the multiplicative group of \v{C}ech r-cochains. Then we also require a nilpotent operator $\delta:C^{r}(M)\rightarrow C^{r+1}(M)$ defined as,
\begin{equation}
 \delta (f(t_0,\ldots,t_r))=\prod^{r+1}_{k=0}f(t_0,\ldots,\hat{t}_k,\ldots,t_{r+1}),
\end{equation}
where the ''hat`` above the k\thth\, entry denotes removal of that quantity. For example consider a 2-cochain $f$. If we allow the boundary operator to act on $f$ we find the following 3-cochain,
\begin{equation}
 \delta (f(t_0,t_1))=f(t_1,t_2)f(t_0,t_2)f(t_1,t_2).
\end{equation}
We can see that $\delta$ is a nilpotent operation, i.e.\ applying $\delta$ twice we will be removing two elements,
\begin{equation}
 \delta(\delta f)=\prod_{j,k=0}^{r+1}f(t_0,\ldots,\hat{t}_k,\ldots,\hat{t}_j,\ldots,t_{r+1}).
\end{equation}
Note that every combination will appear twice since $f$ is invariant under arbitrary permutations,
\begin{align}
 f(t_0,\ldots,\hat{t}_k,&\ldots,\hat{t}_j,\ldots,t_{r+1}) \nonumber\\
 &=f(t_0,\ldots,\hat{t}_j,\ldots,\hat{t}_k,\ldots,t_{r+1}).
\end{align}
Since the values $f$ takes are in $\mathbb{Z}_2$ we have $f^2=1$ and so we see $\delta$ is nilpotent,
\begin{equation}
 \delta^2(f(t_0,\ldots,t_r))=1.
\end{equation}
If we let $Z^r(M;\integers_2)=\{f\in C^r(M)|\delta f=1\}$ be the cocycle group and $B^{r}(M;\integers_2)=\{f\in C^{r}(M)|f=\delta h, h\in C^{r-1}\}$ the coboundary group, we define the \v{C}ech cohomology as,
\begin{equation}
 H^{r}(M;\mathbb{Z}_2)=\frac{Z^r(M;\integers_2)}{B^r(M;\integers_2)}.
\end{equation}
An element $f\in Z^r(M,\integers_2)$ defines a class $[f]\in H^r(M,\integers_2)$ given by the set in \Eref{eq:eqclass},
\begin{equation}
\{f'\in Z^r(M,\integers_2)|f'=\delta h f, h\in C^{r-1}(M,\integers_2)\}.\label{eq:eqclass}
\end{equation}

\section{Cohomology with compact support}\label{app:compact_support}
For a non-compact manifold we instead work with the open sets which cover $M$. In essence we want to define on some compact subset $U\subset M$ a function $\tilde{f}:M\rightarrow \reals$ such that $\tilde{f}|_{U}=f:U\rightarrow\reals$ and $\tilde{f}|_{M-U}=0$, then we say that $f$ has compact support. This idea will be the basis for compactly supported cohomology. We will have cochains that vanish outside of a compact set.

To begin we first define a relative chain group. Suppose we have some topological space $X$ and $A$ a subspace of $X$. Then we have a subset of the chain group $C_n(A)\subset C_n(X)$. The boundary operator $\partial:C_n\rightarrow C_{n-1}$ acting on $C_n(A)$ is the restriction of $\partial$ to $A$ and so gives back $C_{n-1}$. The relative chain group is then given by the quotient $C_n(X,A)=C_n(X)/C_n(A)$. The relative homology group is defined as before as a quotient of the kernel and image of the boundary map $\partial$. The dual system is the relative cohomology $H^r(X,A;G)$ (for a finite abelian group $G$) and is defined as a quotient of the cochain and coboundary group as before only using relative version of the groups instead~\cite{Hatcher2002}. An element in $H^r(X,A;G)$ is a class which vanishes on a set $A\subset X$. We can immediately see the usefulness of this cohomology when the set $A$ is replaced by its complement $X-A$. The cohomology $H^r(X,X-A;G)$ has elements which vanish outside of the set $A$, exactly the condition we need to define cohomology with compact support. 

The final piece we need to build the cohomology with compact support for the composite gauge theory of gravity is a set of compact subsets of the base manifold $M$. Flagga and Antonsen accomplish this with the aide of the paracompactness of $M$ and the fact that every metric space is normal. With these conditions, each locally finite cover $\mathcal{U}$ has a refinement $\mathcal{U'}=\left\{U'|\text{for}\hspace{.1cm}m\in\mathcal{U'},\hspace{.1cm}\cup\{U'\in\mathcal{U'}|m\in \mathcal{U'}\}\subset U\right\}$~\cite{Flagga2004}. A space which meets this criterion is called strongly paracompact~\cite{Flagga2004}. Furthermore if each open cover has a countable subcover then the space is called finally compact and the countable subcover is called a shrinking~\cite{Flagga2004}. With this information we can then find a sequence of compact sets,
\begin{equation}
 k_1\subset \mathring{K}_2 \subset K_2 \subset \cdots \subset\mathring{K}_j \subset K_j,\label{eq:compactsubsets}
\end{equation}
such that $\cup_{j\in \mathbb{N}}\mathring{K}_j=M$ where $\mathring{K}$ denotes the interior of the set K~\cite{Flagga2004}. If we now do as above and create a cochain group $C^n(M,M-K_j,\integers)$, then the compactly supported cochain group is $C^n_c=\cup_{j\in\integers}C^n(M,M-K_j,\integers)$~\cite{Hatcher2002}. Additionally we notice that the coboundary operator moves us within the relative cochain group since if $f\in C^n_c(M;\integers)$ vanishes on $M-K_j$ then so does $\delta f$ and so we can write the cohomology with compact support as,
\begin{equation}
 H^r_c(M;\integers)=\frac{Z^r_c(M;\integers)}{B^r_c(M;\integers)},
\end{equation}
only now we work with the compactly supported cocycle and coboundary groups.

\section*{References}
\bibliographystyle{JHEP}
\bibliography{pubbib}

\end{document}